\documentclass[12pt,preprint]{emulateapj}
\usepackage{natbib}
\usepackage{color}
\usepackage{hyperref} 
\hypersetup{colorlinks=true,citecolor=blue}

\def \bea {\begin{eqnarray}}
\def \ena {\end{eqnarray}}                  
\def    \simlt  {\lower.5ex\hbox{$\; \buildrel < \over \sim \;$}}
\def    \simgt  {\lower.5ex\hbox{$\; \buildrel > \over \sim \;$}}

\newcommand     \mum    {\,\mu{\rm m}}  

\def	\cm		{\,{\rm {cm}}}
\def	\m		{\,{\rm m}}

\def	\B		{{\rm B}}
\def	\D		{{\rm D}}

\def    \exp 		{\,{\rm {exp}}}
\def	\g		{\,{\rm g}}

\def	\GHz		{\,{\rm {GHz}}}

\def	\K		{\,{\rm K}}
\def    \kB    		{k_{\rm B}}

\def	\s		{\,{\rm s}}
\def	\sr		{\,{\rm {sr}}}
\def    \ln  		{\,{\rm {ln}}}
\def    \yr  		{\,{\rm {yr}}}

\def	\H		{\rm H}

\def	\xhat		{\hat{\bf x}}
\def	\yhat		{\hat{\bf y}}
\def	\zhat		{\hat{\bf z}}
\def	\ahat		{\hat{\bf a}}
\def	\ehat		{\hat{\bf e}}

\def    \Bv     	{\bf  B}

\def	\ba			{{\bf a}}

\def	\bB		{{\bf B}}

\def	\bJ		{{\bf J}}

\def	\gas		{\rm {gas}}
\def	\th			{\rm {th}}

\def	\d			{\rm d}

\def	\eff		{\rm {eff}}

\def	\mag 	{\rm {m}}

\def	\tot		{\rm {tot}}

\def	\obs		{\rm {obs}}

\def	\ed			{\rm {ed}}

\font\mib=cmmib10

\def\bmu{\hbox{\mib\char"16}}

\def\bomega{\hbox{\mib\char"21}}

\begin{document}
\shortauthors{Hoang, Vinh, \& Lan}
\title{Spinning dust emission from ultrasmall silicates: Emissivity and Polarization Spectrum}

\author{Thiem Hoang\altaffilmark{1}, Nguyen Anh Vinh\altaffilmark{2}, and Nguyen Quynh Lan\altaffilmark{2}}
\altaffiltext{1}
{Canadian Institute for Theoretical Astrophysics, University of Toronto, 60 St. George Street, Toronto, ON M5S 3H8, Canada}
\altaffiltext{2}
{
Department of Physics, Hanoi National University of Education, 136 Xuan Thuy, Cau Giay, 1000, Hanoi, Vietnam
}

\begin{abstract}
Anomalous microwave emission (AME) is an important Galactic foreground of Cosmic Microwave Background (CMB) radiation. It is believed that the AME arises from rotational emission by spinning polycyclic aromatic hydrocarbons (PAHs) in the interstellar medium (ISM). In this paper, we assume that a population of ultrasmall silicate grains may exist in the ISM, and quantify rotational emissivity from these tiny particles and its polarization spectrum. We found that spinning silicate nanoparticles can produce strong rotational emission when those small grains follow a log-normal size distribution. The polarization fraction of spinning dust emission from tiny silicates increases with decreasing the dipole moment per atom ($\beta$) and can reach $P\sim 20\%$ for $\beta\sim 0.1$D at grain temperature of 60 K. We identify a parameter space $(\beta,Y_{Si})$ for silicate nanoparticles in which its rotational emission can adequately reproduce both the observed AME and the polarization of the AME, without violating the observational constraints by the ultraviolet extinction and polarization of starlight. Our results reveal that rotational emission from spinning silicate may be an important source of the AME.
\end{abstract}


\section{Introduction}

Anomalous microwave emission (AME) in the $\sim$ 10--60 GHz frequency range is a new, important Galactic foreground component, which was discovered about 20 years ago (\citealt{Kogut:1996p5293}; \citealt{Leitch:1997p7359}). Two new emission processes have been proposed to explain the AME, including spinning dust emission (\citealt{1998ApJ...508..157D}, hereafter DL98) from rapidly spinning ultrasmall grains and magnetic dipole emission (hereafter MDE) from magnetic nanoparticles (\citealt{1999ApJ...512..740D}, hereafter DL99).\footnote{In this paper, tiny particles, ultrasmall particles, and nanoparticles are interchangeable, which refers to dust grains of size below $10$nm ($100\AA$).}  

In the spinning dust emission paradigm, the emissivity is determined mostly by three principal parameters, including permanent dipole moment, rotation rate, and the abundance of spinning ultrasmall grains (\citealt{Erickson:1957p4806}; DL98; \citealt{2009MNRAS.395.1055A}; \citealt{Hoang:2010jy}, hereafter HDL10). In that sense, any nanoparticles having permanent dipole moment would produce rotational emission because the nanoparticles of any kind naturally rotate due to gas-grain collisions and absorption of ultraviolet (UV) photons. 

Since the introduction of spinning dust emission, polycyclic aromatic hydrocarbons (hereafter PAHs) are taken to be the unique carrier of rotational emission, because their existence in the interstellar medium was well established (\citealt{1984A&A...137L...5L}; see review by \citealt{2008ARA&A..46..289T}). Theoretical models with spinning PAHs have proved to be successfully in reproducing the observational data from both ground-based experiments \citep{2004ApJ...617..350F} and satellites (\citealt{2011ApJ...741...87H}; \citealt{PlanckCollaboration:2011p515}; \citealt{Collaboration:2013vx}). In particular, \cite{2016MNRAS.456.2290T} recently show that the 1 cm CARMA data from dense cores do not follow the extrapolation of thermal dust radiation and can be fitted with spinning dust emission. 


Aiming to test the hypothesis of rotation emission from spinning PAHs as a source of the AME, \cite{Hensley:2015wca} have carried out an analysis using full-sky AME data from Planck and the all-sky WISE data. Surprisingly, they found no correlation between the PAH abundance and the AME, casting doubt on PAHs as the underlying carrier. It is suggested that the AME may be produced by spinning non-PAH nanoparticles \citep{Hensley:2015wca}, or that the physics of interstellar PAHs is different from the one adopted in spinning dust model \citep{Hoang:2015ts}. This encourages us to search for unexplored possibilities that may be important for the AME.

Recently, \cite{Hoang:2015ts} have quantified rotational emission from spinning iron particles that have intrinsic magnetic dipole moments due to spontaneous magnetization. They found that the rotational emissivity is one order of magnitude lower than that produced by spinning PAHs even when all iron abundance is concentrated within free-flying nanoparticles. Interestingly, their detailed calculations show that the polarization of magnetic dipole emission from free-fliers is low, which suggests a more important role of free-flying nanoparticles as the AME than previously thought. 

The study of this present paper is based on the assumption that there may exist a new population of nanoparticles in the the interstellar medium (ISM), namely silicate nanoparticles, in addition to PAHs and potential iron nanoparticles. Since nano silicates can have permanent electric dipole moments (see Section \ref{sec:rotdamp}), we expect those spinning nanoparticles will produce considerable rotational emission if their abundance is sufficiently large. 

The presence of ultrasmall silicate grains in the ISM remains a hypothesis, in contrast to PAHs where their existence is demonstrated by the 2175$\AA$ extinction bump and mid-infrared (IR) emission features (see \citealt{2011piim.book.....D}). The first estimate by \cite{1986A&A...159..328D} shows that the fraction of total Si abundance (Si/H$= 3.6\times 10^{-5}$) contained in ultrasmall grains, denoted by $Y_{\rm Si}$, is less than $1\%$. However, a later analysis by \cite{Li:2001p4761} shows that $Y_{\rm Si}$ can reach $\sim 10\%$ without violating the observational constraints of the UV starlight extinction and mid-IR emission. If silicate nanoparticles are indeed present in the ISM, their rotational emission will certainly contribute to the AME. The goal of this paper is to employ our advanced spinning model (HDL10; \citealt{2011ApJ...741...87H}, hereafter HLD11) to quantify the rotational emissivity from silicate nanoparticles and to constrain its physical parameters required to reproduce the observed AME in the ISM.

Our paper is structured as follows. In Section \ref{sec:rotdamp} we briefly describe grain properties including electric dipole moments of silicate material. Section \ref{sec:spindust} presents an overview of spinning dust emission mechanism.  Section \ref{sec:num} is devoted to description of rotational damping and excitation of silicate grains, and numerical methods. Results for spinning silicate emission and polarization spectra are presented in Section \ref{sec:spinem}. An extended discussion on implications of rotational emission from spinning silicate for the AME, the AME polarization, and UV polarization of starlight are presented in Section \ref{sec:discus}. The principal results are then summarized in Section \ref{sec:sum}.

\section{Grain Properties of ultrasmall silicates}\label{sec:rotdamp}
\subsection{Grain shape and size}
We consider oblate spheroidal grains with moments of inertia $I_{1}>I_{2}=I_{3}$ along the grain principal axes denoted by $\hat{\ba}_{1}$, $\hat{\ba}_{2}$ and $\hat{\ba}_{3}$. Let $I_{\|}=I_{1}$ and $I_{\perp}=I_{2}=I_{3}$. They take the following forms: 
\bea
I_{\|}=\frac{8\pi}{15}\rho a_{1}a_{2}^{4},~I_{\perp}=\frac{4\pi}{15}\rho a_{2}^{2}a_{1}\left(a_{1}^{2}+a_{2}^{2}\right),\label{eq:Iparperp}
\ena
where $a_{1}$ and $a_{2}=a_{3}$ are the lengths of the semiminor and semimajor axes of the oblate spheroid with axial ratio $s= a_{1}/a_{2}<1$, and $\rho$ is the grain material density. 

The effective grain size $a$ is defined as the radius of an equivalent sphere of the same volume, which is given by
\bea
a=\left(\frac{3}{4\pi} (4\pi/3) a_{1}a_{2}^{2}\right)^{1/3}=a_{2}s^{1/3}.\label{eq:aeff}
\ena

Assuming that all Si, Mg, and Fe abundance is present in dust grains to form the typical structure $Mg_{1.1}Fe_{0.9}SiO_{4}$, the grain material density corresponds to $\rho\approx 4\g\cm^{-3}$. The number of Si atoms is related to the grain size $a$ as $N_{\rm Si}=59.67a_{-7}^{3}$, and the total number of atoms is $N=7N_{\rm Si}=417.74a_{-7}^{3}$.

\subsection{Electric dipole moment}
The rotational emission mechanism is built upon the fact that nanoparticles have non-zero dipole moments. Since the specific composition of ultrasmall silicate is uncertain, its electric dipole moment is rather arbitrary. For instance, a SiO molecule can have a large electric dipole moment of $\mu=3.098\D$, which corresponds to a dipole moment per atom $\beta=1.098\D$. A SiC has a larger dipole moment of $\mu=5.6\D$.  Moreover, the dipole moments of pyroxene structures, enstatite (MgSiO$_{3}$) and ferrosilite (FeSiO$_{3}$), are $\mu=12.2$D and $9.5$D, respectively (see \citealt{2011Icar..212..373S}). Thus, we expect silicate nanoparticles can have quite large dipole moment. Table \ref{tab:dipole} lists the electric dipole moment ($\mu$) of selected bonds and molecules and the dipole per atom ($\beta$). 

Let assume that $N$ atoms in the grain have random distribution, then the intrinsic dipole moment of the grain can be estimated using the random walk for the orientation of individual dipoles:
\bea
\mu_{i}^{2}=N\beta^{2}=66.84(\beta/0.4\D)^{2} a_{-7}^{3} \D^{2}.\label{eq:muin}
\ena

Also, when ultrasmall silicate grains exhibit some asymmetric charge distribution in which the grain charge centroid is displaced from its center of mass, the net dipole moment is non-zero. Thus, the total electric dipole moment of the grain can be written as (DL98):
\bea
\mu^{2}=\mu_{i}^{2}+\epsilon \langle Z^{2}\rangle,\label{eq:musq}
\ena
where $\langle Z^{2}\rangle^{1/2}$ is the rms value of the grain charge, and $\epsilon$ is a parameter taken to be $0.1$ as in DL98.

\begin{table}
\centering
\caption{Electric dipole moments\\
 of selected molecules}\label{tab:dipole}
\begin{tabular}{lllll} \hline\hline\\
{\it Bond} & \multicolumn{1}{c}{$\mu$ (D)} & \multicolumn{1}{c}{\it Molecule} & \multicolumn{1}{c}{$\mu$ (D)} & \multicolumn{1}{c}{$\beta$ (D)}\\[1mm]
\hline\\
Si-H & 1.0 & SiO &3.098 & 1.549 \\[1mm]
Si-C &1.2 & SiC & 5.6 & 2.8 \\[1mm]
Si-N & 1.55 & SiC$_{2}$ &2.393 & 0.797 \\[1mm]
&  & MgO &$6.2$ & 2.6 \\[1mm]
&& MgSiO$_{3}$ & 12.2 & 2.44\\[1mm]
&& FeSiO$_{3}$ &9.5 & 1.9 \\[1mm]

\\[1mm]
\hline\hline\\
\end{tabular}
\end{table}

\section{Overview of Spinning Dust Emission Mechanism from Wobbling Grains}\label{sec:spindust}
Consider a grain with dipole moment $\bmu$ fixed in the grain body rotating with angular momentum $\bJ$. If the grain only spins around its symmetry axis $\ahat_{1}$ ($\ahat_{1}\|\bJ$), then the spinning dipole moment emits radiation at a unique frequency $\nu$ equal to the rotational frequency, i.e., $\nu=\omega/2\pi$ (DL98). The power emission by the spinning grain at frequency $\nu$ is
\bea
P_{\ed}=\left(2\mu_{\perp}^{2}\omega^{4} \over 3c^{3}\right)~~~,\label{eq:Ped}
\ena
where $\mu_{\perp}$ is the component of $\bmu$ perpendicular to the rotation axis.

For more realistic rotation with $\ahat_{1}$ misaligned from $\bJ$ (HDL10; HLD11), an isolated grain of conserved $\bJ$ essentially emits radiation at four frequency modes, which are a function of $J$ and the angle $\theta$ between $\ahat_{1}$ and $\bJ$. Following HLD11, the four frequency modes are described by
\bea
\omega_{m_{i}}=\Omega(1+i(h-1)\cos\theta),\label{eq:omegak}\\
\omega_{n_{1}}=\Omega(1-h)\cos\theta,
\ena
with $i=0,\pm1$, $h=I_{\|}/I_{\perp}$, and $\Omega=J/I_{\|}$. 

Due to thermal fluctuations within the grain, the angle $\theta$ changes rapidly, which results in the wobbling of the grain with respect to $\bJ$ (see the next section). For the case of efficient internal relaxation (see HLD11 for details), the thermal fluctuations of $\theta$ can be described by a distribution function:
\bea
f_{\rm LTE}(\theta,J)\propto \exp\left(\frac{J^{2}}{2I_{\|}kT_{d}}\left[1+(h-1)\sin\theta^{2} \right] \right)\sin\theta,\label{eq:fLTE}
\ena
where $T_{d}$ is the dust grain temperature, and $f_{\rm LTE}(\theta,J)d\theta$ describes the probability of finding $\theta$ between $\theta, \theta+d\theta$ for given $J$.

The grain angular momentum changes randomly due to a variety of interaction processes (see DL98) and is described by a distribution function $f_{J}(J)$, for which $f_{J}(J)dJ$ indicates the probability of finding the grain angular momentum between $J, J+dJ$. 

The rotational emissivity at observation frequency $\nu$ ($\omega/(2\pi)$) from a grain of size $a$ is obtained by integrating the power emission over the angular momentum distribution and summing over all emission modes:
\bea
j_{\nu}^{a}&\equiv&\frac{1}{2} \frac{f_{J}(I_{\|}\omega/h)}{h}
\frac{2\mu_{\|}^{2}}{3c^{3}}\omega^{4}\langle\sin^{2}\theta\rangle\nonumber\\
&&+\frac{1}{2} \frac{\mu_{\perp}^{2}}{6c^{3}}\omega^{4}\int_{J_{l}}^{J_{u}}
(1+\cos\theta_{m_{+1}})^{2}pdf_{\m_{1}}(\omega|J)
f_{J}(J)dJ\nonumber\\
&&+\frac{1}{2} \frac{\mu_{\perp}^{2}}{6c^{3}}\omega^{4}\int_{J_{l}}^{J_{u}}
(1-\cos\theta_{m_{-1}})^{2}pdf_{\m_{-1}}(\omega|J)
f_{J}(J)dJ\nonumber\\
&&+\frac{1}{2} \frac{\mu_{\perp}^{2}}{3c^{3}}\omega^{4}\int_{J_{l}}^{J_{u}}
\sin^{2}\theta pdf_{n_{1}}(\omega|J)
f_{J}(J)dJ,\label{eq:jnua}
\ena
where $\mu_{\|}=\mu \cos\theta$, $\mu_{\perp}=\mu \cos\theta$, the lower and upper limits of $J$ are determined by $\cos\theta=-1$ and $1$ from setting $\omega_{k}$ equal to the observed frequency $\omega$, $\cos\theta_{m_{+1}}=(-\omega/\Omega +h)/(h-1)$, $\cos\theta_{m_{-1}}=(\omega/\Omega -h)/(h-1)$ (using Eqs. \ref{eq:omegak}), and $pdf_{k}(\omega|J)$ is the probability of finding the emission at frequency $\omega$ by mode $\omega_{k}$.

It is straightforward to obtain $J_{l}=I_{\|}\omega/(2h-1)$ and $J_{u}=I_{\|}\omega$ for $m_{+1}$ mode, $J_{u}=I_{\|}\omega/(2h-1)$ and $J_{l}=I_{\|}\omega$ for $m_{-1}$ mode, and $J_{l}=I_{\|}\omega/(h-1)$ and $J_{u}=\infty$ for $n_{1}$ mode.

The probability $pdf_{k}(\omega|J)$ is given by (HLD11):
\bea
pdf_{m_{\pm 1}}=f_{\rm LTE}(\theta,J)\frac{\pm 1}{\Omega (h-1)},\\
pdf_{n_{1}}=f_{\rm LTE}(\theta,J)\frac{1}{\Omega (h-1)},
\ena
where $f_{\rm LTE}$ is given by Equation (\ref{eq:fLTE}). It is easy to see that the second and third term in Equation \ref{eq:jnua} is the same because $ (1+\cos\theta_{m_{+1}})^{2}=(1-\cos\theta_{m_{-1}})^{2}$.

The rotational emissivity from all spinning ultrasmall grains is calculated by integrating over the grain size distribution $dn/da$:
\bea
j_{\nu}&=&\int_{a_{\min}}^{a_{\max}} j_{\nu}(a) \frac{n_{\H}^{-1}dn}{da}da, \label{eq:jnu}\\
\ena
where $a_{\min}=0.35$nm and $a_{\max}=10$nm are assumed, as in previous works (DL98; HDL10). 

Therefore, the calculations of rotational emissivity depend essentially on the dipole moment $\mu$, the distribution function $f_{J}$, and $dn/da$. In the following, we will discuss a bit more detail for spinning silicate grains.

\section{Rotational Dynamics and Numerical Method}\label{sec:num}
The discussion on rotational dynamics of silicate nanoparticles in this section is in analogy to that of spinning PAHs, which is presented in detail in DL98 and HDL10. Below, we provide a brief description for reference.
\subsection{Rotational Damping and Excitation Coefficients}

Rotational damping and excitation for dust grains, in general, arise from collisions between grains and gaseous atoms followed by the evaporation of atoms/molecules from the grain surface, absorption of starlight and IR emission (DL98; HDL10). If the grain possesses electric dipole moment, the distant interaction of the grain electric dipole with passing ions results in an additional effect, namely plasma drag. The rotational damping and excitation for these processes are described by the dimensionless damping coefficient $F$ and excitation coefficient $G$, respectively (see HDL10 for more detail). 

For calculations of the $F$ and $G$ coefficients, one needs to know the charge distribution $f_Z$ of nano silicates. As in \cite{2012ApJ...761...96H}, we find $f_Z$ by solving the ionization equilibrium equations that takes into account collisional charging \citep{1987ApJ...320..803D} by sticking collisions with electrons and ions and charging by photoemission \citep{2001ApJS..134..263W}. Dielectric function of astronomical silicate is adopted for nano silicates.

Moreover, rotational emission by electric dipole moment results in the rotational damping. The characteristic damping time due to dipole emission $\tau_{\ed}$ depends on the dipole moment $\beta^{2}$ (see Appendix \ref{apdx:FGcoeff}). We assume that the electric dipole is assumed to be in the plane perpendicular to the grain symmetry axis. It is noted that, as found in HDL10, the difference with the result from the case of isotropic orientation is small, within $10\%$, because strong internal fluctuations average out the dipole orientation with $\bJ$.

\begin{figure}
\includegraphics[width=0.4\textwidth]{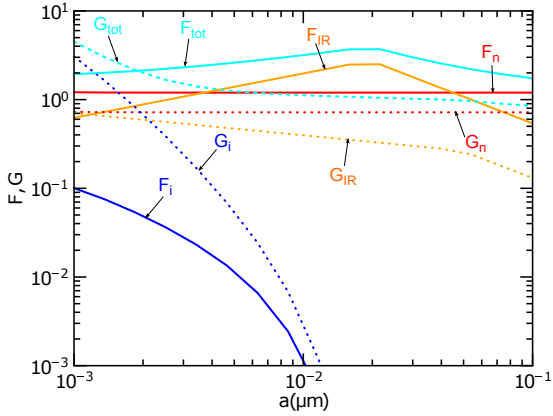}
\caption{Rotational damping ($F_{j}$; solid lines) and excitation ($G_{j}$; dotted lines) coefficients where $j=i, n, IR$ denotes the collision with incident ions, incident neutrals, and IR emission for silicate nanoparticles in the CNM.}
\label{fig:FG}
\end{figure}

Figure \ref{fig:FG} shows the obtained values of $F$ and $G$ for various processes for silicate grains in the cold neutral medium (CNM; gas density $n_{\H}=30\cm^{-3}$, ionization fraction  $x_{\H}=10^{-3}\cm^{-3}$, and temperature $T_{\gas}=100\K$) computed for an oblate spheroidal grain rotating along its symmetry axis. As shown, IR emission is dominant for rotational damping of small grains ($a<0.05\mum$), and neutral collisions are dominant for rotational damping/excitation for larger grains. Ion collisions dominate rotational excitation for ultrasmall grains because such grains are negatively charged due to sticking collisions with electrons that outperform photoelectric emission.

\subsection{Numerical Method: Langevin Equations}
Following our previous works (HDL10, HLD11), to find the accurate distribution of grain angular momentum, we numerically solve the Langevin equations for the evolution of $\bJ$ in time in an inertial coordinate system, denoted by unit vectors $\ehat_{1},~\ehat_{2},$ and $\ehat_{3}$, where $\ehat_{1}$ is chosen to be parallel to the ambient magnetic field $\Bv$ (see Figure \ref{fig:RFs}). The Langevin equations read 
\bea
dJ_{i}=A_{i}dt+\sqrt{B_{ii}}dW_{i}\mbox{~for~} i=~1,~2,~3,\label{eq:dJ_dt}
\ena
where $dW_{i}$ are the random variables drawn from a normal distribution with zero mean and variance $\langle dW_{i}^{2}\rangle=dt$, and $A_{i}=\langle {\Delta J_{i}}/{\Delta t}\rangle$ and $B_{ii}=\langle \left({\Delta J_{i}}\right)^{2}/{\Delta t}\rangle$ are the damping and excitation coefficients defined in the $(\ehat_{1},\ehat_{2},\ehat_{3})$ frame. 

The damping and excitation coefficients in a frame of reference fixed to the grain body, $A_{i}^{b}$ and $B_{ij}^{b}$, are related to the dimensionless damping and excitation coefficients ($F$ and $G$) as follows:
\bea
A_{i}^{b}&=&-\frac{J_{i}^{b}}{\tau_{\gas,i}}=-\frac{J_{i}^{b}}{\tau_{\H,i}}F_{\tot,i},\\
B_{11}^{b}&=&B_{\|}=\frac{2I_{\|}\kB T_{\rm gas}}{\tau_{\rm H,\|}}G_{\rm tot,\|},\\
B_{22}^{b}&=&B_{33}^{b}=B_{\perp}=\frac{2I_{\perp}\kB T_{\rm gas}}{\tau_{\rm H,\perp}}G_{\rm tot,\perp},
\ena
where $\tau_{H,\|},\tau_{\H,\perp}$ are rotational damping due to gas collisions for rotation parallel and perpendicular to the symmetry axis (see Appendix \ref{apdx:FGcoeff}), $F_{\tot,i}$ and $G_{\tot, ii}$ for $i=1,2,3$ (or $\|, \perp$) are the total damping and excitation coefficients from various processes, and $\tau_{\gas, i}=\tau_{\H,i}/F_{\tot,i}$.  Finally, $A_{i}$ and $B_{ii}$ are obtained by using the transformation of coordinate systems for $A_{i}^{b}, B_{ii}^{b}$ from the $(\ahat_{1},\ahat_{2},\ahat_{3}$) frame to  the $(\ehat_{1},\ehat_{2},\ehat_{3}$) frame (see HLD10; \citealt{2014ApJ...790....6H}).

To study the effect of magnetic relaxation on the alignment of nanoparticles, we incorporate a damping term $-{J_{2,3}}/{\tau_{\mag}}$ to $A_{2,3}$ and an excitation term $B_{{\mag},22}=B_{{\mag},33}=\delta_{m}T_{\gas}/T_{d}$ to $B_{22}$ and $B_{33}$, respectively (see Appendix \ref{sec:DG}). In the dimensionless units of $J'\equiv J/I_{\|}\omega_{T}$ and $t'\equiv t/\tau_{\H,\|}$ with $\omega_{T}$ being thermal angular velocity (Equation \ref{eq:omegaT}), Equation (\ref{eq:dJ_dt}) becomes 
\bea
dJ'_{i}=A'_{i}dt'+\sqrt{B'_{ii}}dw'_{i} \mbox{~for~} i= 1,~2,~3,\label{eq:dJp_dt}
\ena
where $\langle dw_{i}^{'2}\rangle=dt'$ and
\bea
A'_{i}&=&-{J'_{i}}\left[\frac{1}{\tau'_{\gas,{\eff}}} +\delta_{m}(1-\delta_{1i})\right] -\frac{2}{3}\frac{J_{i}^{'3}}
{\tau'_{\ed,{\eff}}},\label{eq:Ai}~~~~\\
B'_{ii}&=&\frac{B_{ii}}{2I_{\|}\kB T_{\gas}}\tau_{\H,\|}+\frac{T_{\d}}{T_{\gas}}\delta_{\mag}(1-\delta_{1i}).\label{eq:Bii}
\ena
Above, $\delta_{1i}=1$ for $i=1$ and $\delta_{1i}=0$ for $i\ne 1$, and
\bea
\tau'_{\gas,{\eff}}= \frac{\tau_{\gas,{\eff}}}{\tau_{\H,\|}},~\tau'_{\ed,{\eff}}&=&\frac{\tau_{\ed,{\eff}}}{\tau_{\H,\|}},~~~
\ena
where $\tau_{\gas,{\eff}}$ and $\tau_{\ed,{\eff}}$ are the effective damping times due to dust-gas interactions and electric dipole emission (see Eq. E4 in HDL10). 

To numerically solve the Langevin equation (\ref{eq:dJp_dt}), we use the second-order integrator, for which the angular momentum component $j_{i}\equiv J'_{i}$ at iterative step $n+1$ is evaluated as follows:
\bea
{j}_{i;n+1} &=& j_{i;n}  -  \gamma_{i}{j}_{i;n}h+\sqrt{h}\sigma_{ii}{\xi}_{n}- \gamma_{i}\mathcal{A}_{i;n}-\gamma_{\ed}\mathcal{B}_{i;n},~~~
\ena
where $h$ is the timestep, $\gamma_{i}=1/\tau'_{\gas,{\eff}} +\delta_{m}(1-\delta_{zi})$, $\gamma_{\ed}=2/(3\tau'_{\ed,\eff})$, $\sigma_{ii}=\sqrt{B'_{ii}}$, and
\bea
\mathcal{A}_{i;n}&=& -\frac{h^{2}}{2} \gamma_{i} j_{i;n}+\sigma_{ii} h^{3/2}g(\xi_{n},\eta_{n})-\gamma_{\ed}j_{i;n}^{3}\frac{h^{2}}{2},\\
\mathcal{B}_{i;n}&=&j_{i;n}^{3}h - 3\gamma_{i}j_{i;n}^{3}\frac{h^{2}}{2}-\frac{3j_{i;n}^{5}\gamma_{\ed}h^{2}}{2}+3j_{i;n}^{2}\sigma_{ii} h^{3/2}g(\xi_{n},\eta_{n}),
\ena
with ${\eta}_{n}$ and ${\xi}_{n}$ being independent Gaussian variables with zero mean and unit variance and $g(\xi_{n},\eta_{n})= \xi_{n}/2 + \eta_{n}/2\sqrt{3}$ (see \citealt{Hoang:2015ts} for details).

For silicate nanoparticles and PAHs with efficient rotational emission, the timestep $h$ is determined essentially by the two timescales, the shortest one $\tau_{\rm ed}$ and the long one $\tau_{\gas}$, which is chosen as $h= 0.01\min[1/F_{\tot,\|}, 1/G_{\tot,\|}, \tau_{\rm ed,\|}/\tau_{\H,\|},1/\delta_{m}]$.  A fixed integration time $T=10^{3}\tau_{\gas}$ is chosen, which ensures that $T$ is much larger than the longest dynamical timescale to provide good statistical calculations of the degrees of grain alignment.  As usual, the initial grain angular momentum is assumed to have random orientation in the space and magnitude $J=I_{\|}\omega_{T}$ (i.e., $j=1$). The solutions $J_{x},J_{y},J_{z}$ at each timestep are tabulated to find the grain angular momentum distribution function $f_{J}$ and degree of alignment (see the next section).

\subsection{Degrees of Grain Alignment}
Let $Q_{X}=\langle G_{X}\rangle $ with $G_{X}=\left(3\cos^{2}\theta-1\right)/2$ be the degree of internal alignment of the grain symmetry axis $\ahat_{1}$ with $\bJ$, and let $Q_{J}=\langle G_{J}\rangle$ with $G_{J}=\left(3\cos^{2}\zeta-1\right)/2$ be the degree of external alignment of $\bJ$ with $\Bv$. Here $\theta$ is the angle between $\hat{\ba}_{1}$ and $\bJ$, $\zeta$ is the angle between $\bJ$ and $\Bv$ (see Figure \ref{fig:RFs}). The angle brackets denote the average over the ensemble of grains. The net degree of alignment of $\ahat_{1}$ with $\Bv$, namely the Rayleigh reduction factor, is defined as $R = \langle G_{X}G_{J}\rangle$.

The angular momentum $\bJ$ and the angle $\zeta$ between $\bJ$ and $\Bv$ obtained from the Langevin equations are employed to compute the degrees of alignment, $Q_{J}$ and $R$. For instance, the Raleigh reduction factor $R$ is calculated as follows: 
\bea
 R\equiv \sum_{n=0}^{N_{\rm step}-1} \frac{G_{X}(\cos^{2}\theta)G_{J}(\cos^{2}\zeta_{n})}{N_{\rm step}},
 \ena
where $G_{X}$ can be replaced by $q_{X} (J_{n})=\int_{0}^{\pi}G_{X}(\cos^{2}\theta)f_{\rm LTE}(\theta, J_{n})d\theta$  in the case of fast internal relaxation (see e.g, RL99). 

\section{Rotational emissivity from spinning tiny sillicates and Polarization}\label{sec:spinem}
\subsection{Grain size distribution}
Following \cite{Li:2001p4761}, we assume that ultrasmall silicate grains follow a log-normal size distribution:
\bea
\frac{1}{n_{\H}}\frac{dn}{da} = \frac{B}{a}\exp\left(-0.5\left[\frac{\log (a/a_{0})}{\sigma}\right]^{2}\right) ,\label{eq:dnda_log}
\ena
where $a_{0}$ and $\sigma$ are the model parameters, and $B$ is a constant determined by $Y_{\rm Si}$.

The peak of the mass distribution $a^{3}dn/d\ln~a$ occurs at $a_{p}=a_{0}e^{3\sigma^{2}}$. For different models shown in Table 1 of \cite{Li:2001p4761} with $a_{0}=3-6\AA$ and $\sigma=0.3-0.6$, the upper limits for $Y_{\rm Si}$ for amorphous ultrasmall silicate grains is $Y_{\rm Si}^{\max}=20, 20,20, 30\%$ (\citealt{Li:2001p4761}).

\begin{figure}
\includegraphics[width=0.45\textwidth]{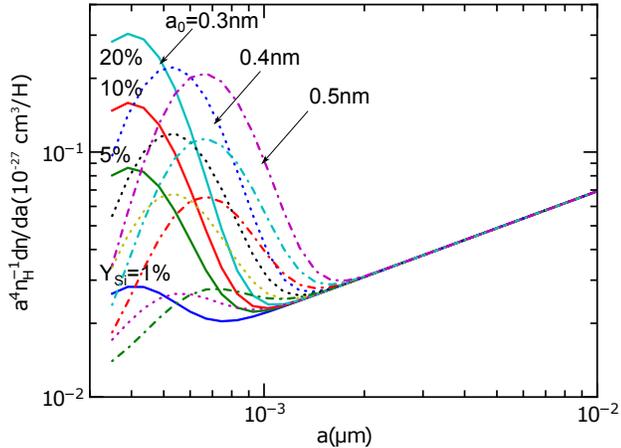}
\caption{Various log-normal size distributions of ultrasmall silicate grains for $Y_{\rm Si}=1,5,10,20\%$, $a_{0}=0.3, 0.4, 0.5$nm and $\sigma=0.3$.}
\label{fig:dnda}
\end{figure}

Figure \ref{fig:dnda} shows the size distribution for different model parameters with $Y_{\rm Si}=1-20\%$, where we have added the power law term $A_{\rm MRN}a^{-3.5}$ (\citealt{Mathis:1977p3072}) with $A_{\rm MRN}=10^{-25.11}\cm^{-2.5}$ for silicate grains larger than $20$nm, although the latter, slowly rotating grains, mostly do not contribute to rotational emission.

\subsection{Rotational Emissivity and Polarization}
Using the angular momentum distribution obtained from Langevin equation and $dn/da$ from Equation (\ref{eq:dnda_log}), we can compute rotational emissivity by spinning silicate using Equation (\ref{eq:jnu}). We consider oblate spheroidal grains with axial ratio $r=2$ for a typical model. To explore the importance of spinning silicate, we consider a model of the ISM with typical physical parameters characterized by the CNM and adopt a typical magnetic field strength $B=10\mu$G. 

In addition to the rotational emissivity computed by Equation (\ref{eq:jnu}), we are also interested in the polarized emissivity, which can be calculated as:
\bea
q_{\nu}&=&\int_{a_{\min}}^{a_{\max}} Q_{J}(a)\cos^{2}\gamma_{B} j_{\nu}(a) \frac{n_{\H}^{-1}dn}{da} da,\label{eq:jpol}
\ena
where $Q_{J}$ is the degree of alignment of grain size $a$, and $\gamma_{B}$ is the angle between $\Bv$ and the plane of the sky. 
 
\begin{figure*}
\centering
\includegraphics[width=0.8\textwidth]{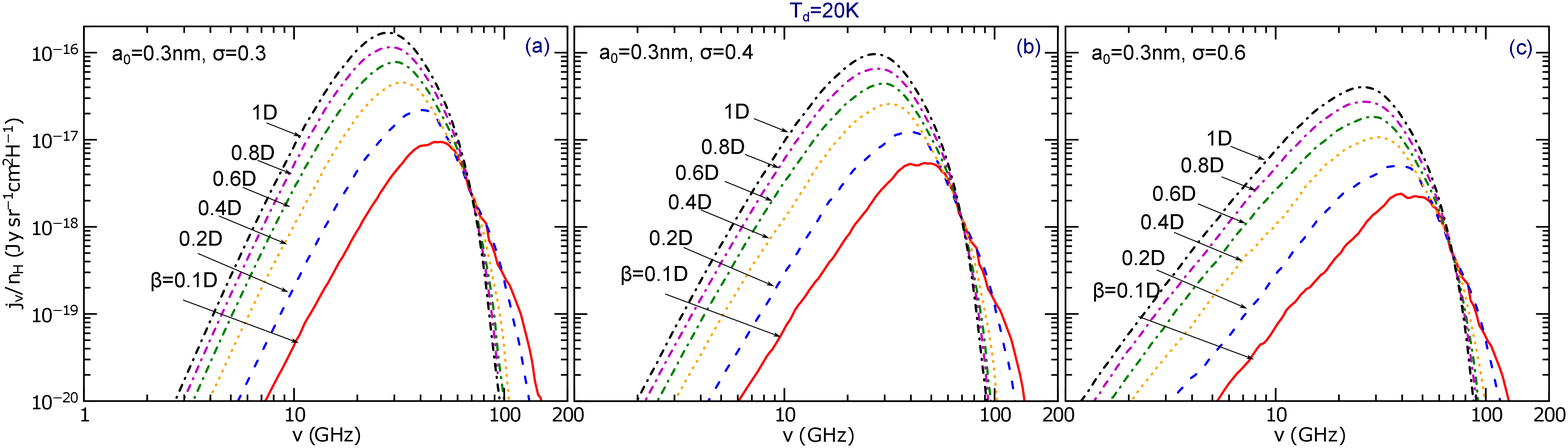}
\includegraphics[width=0.8\textwidth]{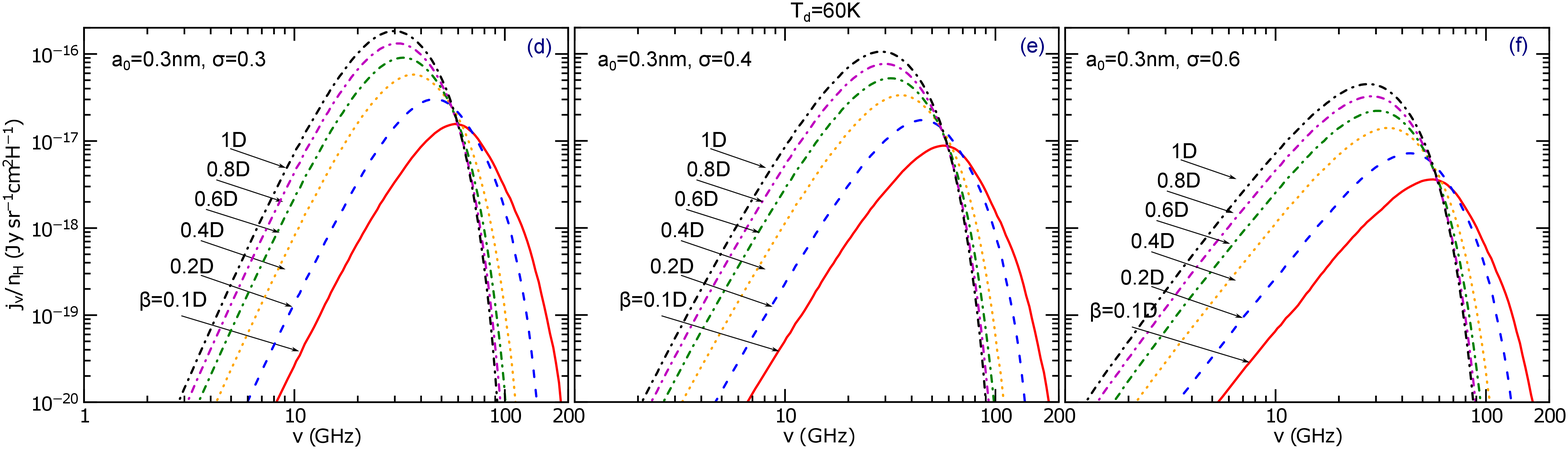}
\caption{Spinning dipole emissivity by ultrasmall silicates computed for the different values of $\beta$. Results for $\sigma=0.3, 0.4, 0.6$ are shown in ((a), (d)), ((b),(e)), and ((c),(f)), respectively. Two temperatures $T_{d}=20$K (upper panels) and 60K (lower panels) are considered.}
\label{fig:jnu_sil}
\end{figure*}

Figure \ref{fig:jnu_sil} shows spinning dust emissivity for the different size distribution parameters with $Y_{\rm Si}=Y_{\rm Si}^{\max}$. Here we take $Y_{\rm Si}^{\max}=20, 20, 30\%$ for $\sigma=0.3, 0.6$ and $0.6$ estimated from \cite{Li:2001p4761}. A range of the dipole moment $\beta$ from $0.1$ to $1$D is considered. As $\beta$ increases, the peak emissivity increases while the peak frequency decreases as a result of the stronger rotational damping by electric dipole emission. In addition, for a fixed $a_{0}=0.3$nm, the peak emissivity decreases as $\sigma$ increases from $0.3$ to $0.6$ due to the increase in the peak size of nanoparticles $a_{p}$. The peak emissivity is slightly larger for the case $T_{d}=60$K due to the additional excitation arising from magnetic fluctuations (see Appendix \ref{sec:DG}).

\begin{figure*}
\centering
\includegraphics[width=0.8\textwidth]{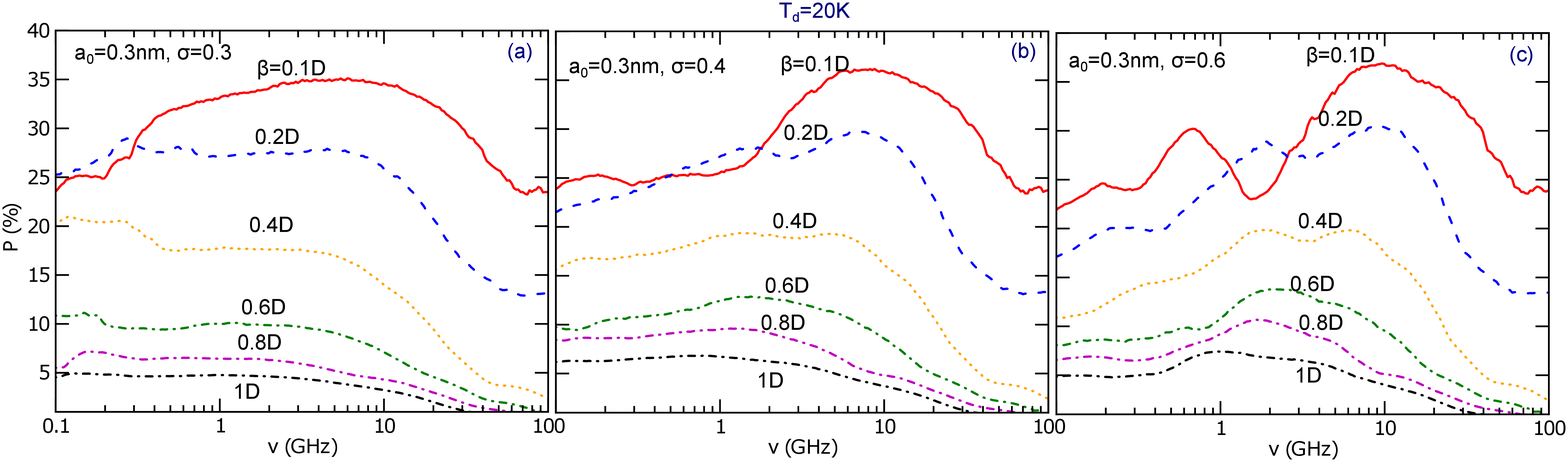}
\includegraphics[width=0.8\textwidth]{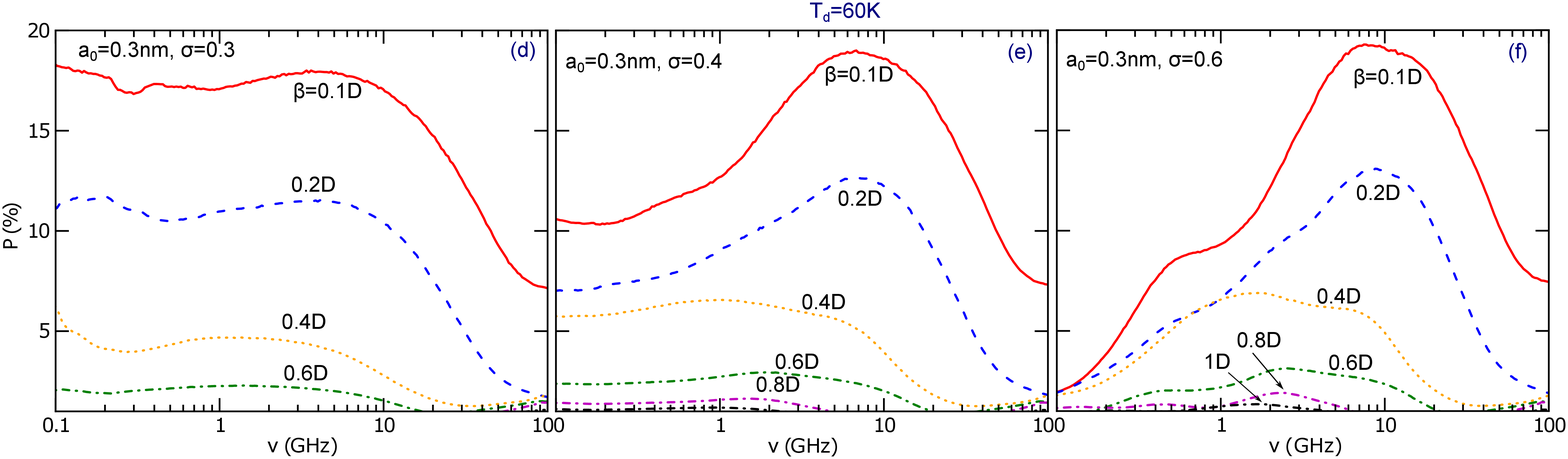}
\caption{Similar to Figure \ref{fig:jnu_sil} but for the polarization spectra of rotational emission. The polarization fraction varies significantly with the values $\beta$.}
\label{fig:pol_sil}
\end{figure*}

Figure \ref{fig:pol_sil} shows the polarization fraction of the rotational emission ($p(\nu)=q_{\nu}/j_{\nu}$) for similar realizations in Figure \ref{fig:jnu_sil}. The magnetic field direction is assumed in the plane of the sky ($\gamma_{B}=0$). Colder spinning silicates tend to generate stronger polarized emission due to weaker internal thermal fluctuations.  Interestingly, we can see that the larger dipole $\beta$ results in a small polarization fraction. This is because the larger dipole induces faster electric dipole damping. This lowers substantially the rms angular momentum of grains and reduces the degree of grain alignment, according to the magnetic alignment mechanism. For considerable values of $\beta\ge 0.4$D, the peak polarization fraction is essentially below $10\%$. However, for $\beta<0.4$D, the peak polarization can reach $5-15\%$ for $\nu=20-60$GHZ.

\section{Discussion}\label{sec:discus}
\subsection{Can rotational emission from spinning silicate reproduce the AME?}
Spinning dust emission is a new mechanism to produce microwave rotational emission, which essentially depends on the permanent dipole moment, rotation rate, and the abundance of ultrasmall grains (DL98; HDL10). In that sense, any nanoparticles owning a permanent dipole moment will produce rotational emission, because nanoparticles of any kind always spin due to gas-grain collisions and UV photon absorption. Previous works have quantified rotational emission by spinning PAHs (DL98; HDL10; HLD11) and spinning iron nanoparticles \citep{Hoang:2015ts}. In this paper, we have quantified rotational emission from a potential dust population, so-called silicate nanoparticles. 

Assuming a log-normal size distribution for silicate nanoparticles with maximum Si abundance contained in ultrasmall grains $Y_{\rm Si}^{\max}=20\%$, we find that the spinning dust emissivity can span $j_{\nu}/n_{\H}\sim 10^{-18}-10^{-16}Jy\cm^{2}/\sr/\H$, which can even exceed the AME level of $j_{\obs}\sim 10^{-17} Jy cm^{2}/\sr/\H$ from the ISM. 

\begin{figure*}
\centering
\includegraphics[width=0.8\textwidth]{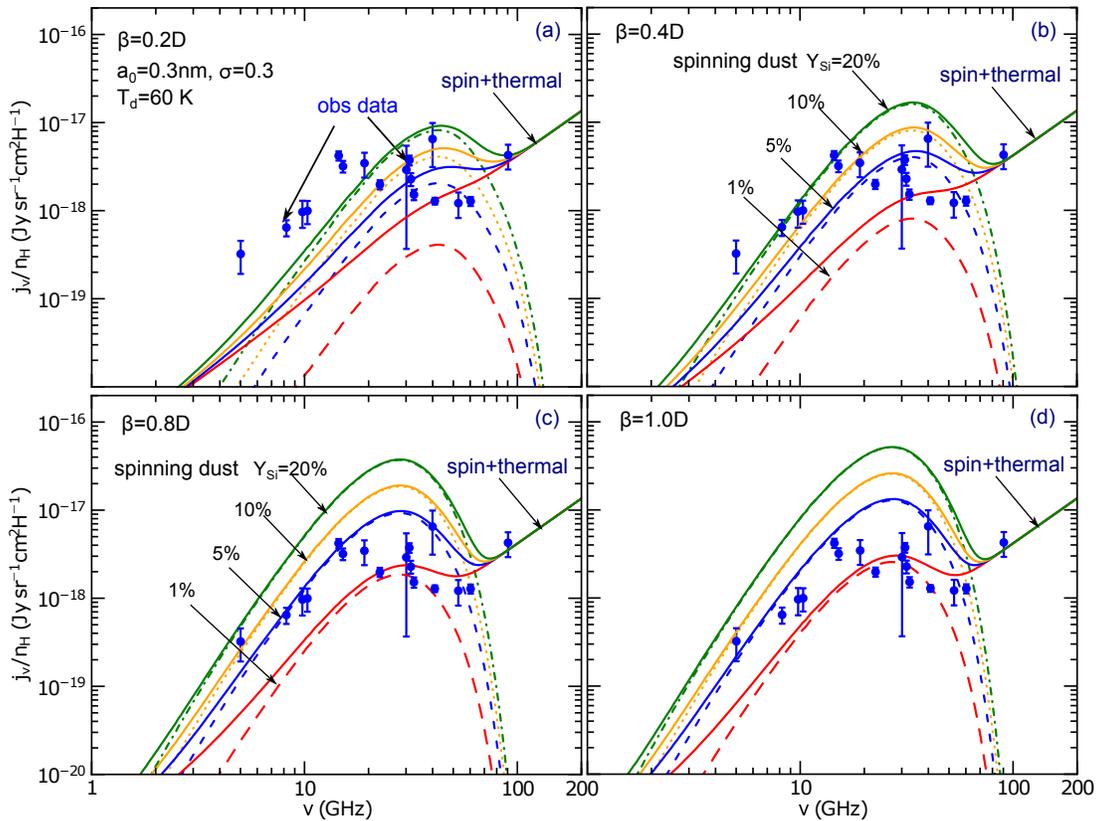}
\caption{Rotational emissivity for $Y_{\rm Si}=1, 5, 10$ and 20$\%$ (dashed and dashed-dotted lines), with a chosen grain size distribution of $a_{0}=0.4nm,\sigma=0.4$. Solid lines show the total emissivity from spinning dust and thermal dust (see text). Different values of $\beta=0.2, 0.4, 0.8, 1D$ are considered. Observational data from various instruments extracted from \cite{2004ApJ...617..350F} are shown in symbols.}
\label{fig:jnu_sil2}
\end{figure*}

To see what value of $Y_{\rm Si}$ the spinning nano silicate can reproduce the AME, in Figure \ref{fig:jnu_sil2} we plot $j_{\nu}$ calculated for $Y_{\rm Si}=1, 5, 10, 20\%$, and selected $\beta=0.2, 0.4, 0.8, 1$D. Since nanoparticles are hotter than big grains, we take $T_{d}=60$K as in previous works (HLD10; HLD11). The observational data from various instruments extracted from \cite{2004ApJ...617..350F} are shown in symbols, and the solid lines are the total spinning dust plus thermal dust emissivity described by a modified black body $j_{\nu}^{\rm td}=j_{94\GHz}(\nu/94.)^{-1.7}$ with $j_{94\GHz}=0.8$MJy/sr. It can be seen that spinning nanosilicate with $Y_{Si}<10\%$ and $\beta>0.2$D (Figure 5(b), (c), (d)) can successfully reproduce the observed AME, in terms of both peak emissivity and frequency. Nano silicate with larger value $\beta$ requires lower abundance $Y_{Si}$ to reproduce the AME.

It is noted that both Wilkinson Microwave Anisotropy Probe (WMAP) and Planck all-sky fitting to the AME requires two spinning dust components, one with a low peak frequency and the other with a higher peak frequency \citep{2015A&A...576A.107P}. In HLD11, it is successfully reproduced by rotational emission from spinning PAHs from the CNM and warn ionized medium (see also \citealt{2015A&A...576A.107P}). In the light of this study, two components may arise from spinning dust in the same phase but with two separate populations, namely spinning PAHs and spinning ultrasmall silicate. In a future paper, we will explore the fitting rotational emission from spinning PAH and spinning silicate to the AME obtained from Planck data to explain two spinning dust components.

\subsection{Constraints from the AME polarization}
\begin{figure}
\includegraphics[width=0.45\textwidth]{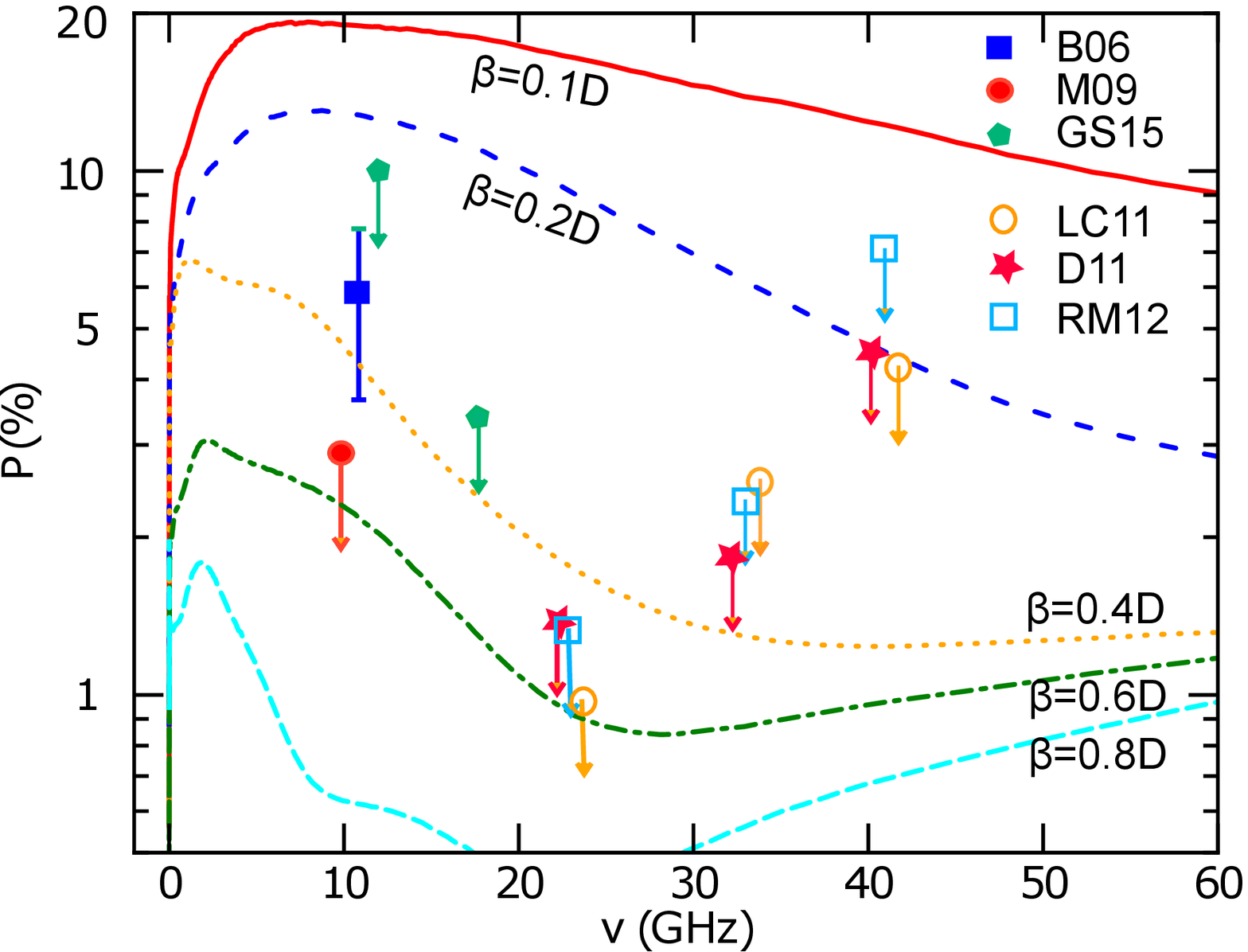}
\caption{Polarization spectrum of spinning silicate emission for the different values of $\beta$. Observational data for the AME polarization from \cite{2006ApJ...645L.141B} (B06), \cite{2009ApJ...697.1187M} (M09), \cite{2011MNRAS.418L..35D} (D11), \cite{LopezCaraballo:2011p508} (LC11), \cite{RubinoMartin:2012ji} (RM12), and \cite{GenovaSantos:2015hc} (GS15), are shown in symbols, where downward arrows indicate the upper limits. The theoretical polarization mostly exceeds the  observed AME polarization for $\beta<0.2$D.}
\label{fig:pol_beta}
\end{figure}

We have found that the polarization fraction of rotational emission by ultrasmall silicate grains varies significantly with electric dipole moment $\beta$, and increases with decreasing $\beta$. Thus, the AME polarization can be used to constrain the value of $\beta$ because the polarization fraction is independent on $Y_{\rm Si}$. 

Figure \ref{fig:pol_beta} shows the polarization fraction for the different $\beta$ overplotted with the AME polarization data observed by various experiments. As shown, the predicted polarization for spinning silicate with $\beta<0.2$D exceeds most of the observed data, whereas spinning silicate with $\beta\ge 0.2$D can adequately reproduce the level of the observed AME polarization.

It is also cautious that the polarization of magnetic dipole emission from free-flying iron nanoparticles is found to be within the AME polarization, therefore, the constraint by the AME polarization is not so restrictive.

\subsection{Constraints from the UV polarization of starlight}
Interstellar polarization of starlight was discovered more than 60 years ago (\citealt{Hall:1949p5890}; \citealt{Hiltner:1949p5856}). It now becomes established that the interstellar polarization is produced by aspherical grains aligned with the magnetic field (see \citealt{LAH15} for a review). For sub-micron grains, the leading mechanism of grain alignment is radiative torque alignment, which was first introduced by \cite{1976Ap&SS..43..291D}, numerically demonstrated in \cite{1996ApJ...470..551D}, and analytically modeled by \cite{2007MNRAS.378..910L}. For nanoparticles, paramagnetic relaxation (\citealt{1951ApJ...114..206D}; \citealt{Jones:1967p2924}; \citealt{1999MNRAS.305..615R}) can induce weak alignment \citep{2014ApJ...790....6H}. It is quantified in \cite{2014ApJ...790....6H} that small grains dominate the UV starlight polarization. Therefore, the presence of interstellar silicate nanoparticles will obviously affect the UV polarization. 

To see for what values of $Y_{Si}$ and $\beta$, the predicted polarization by silicate nanoparticles do not violate the observed UV polarization, we compute the polarization curves of starlight for different values of $Y_{Si}$ and $\beta$. To this end, we first adopt the best-fit model for the ISM  from \cite{2014ApJ...790....6H} that reproduces the "observed" polarization curve described by the Serkowski law (\citealt{Serkowski:1975p6681}) with $\lambda_{\max}=0.55\mum$ and extinction curve. Then, we introduce a population of silicate nanoparticles with the size distribution (Eq. \ref{eq:dnda_log}) and the alignment degree computed in Section \ref{sec:num}. The final polarization curve is then computed using Equation (14) in \cite{2013ApJ...779..152H} (see also \citealt{2014ApJ...790....6H}).

\begin{figure}
\includegraphics[width=0.45\textwidth]{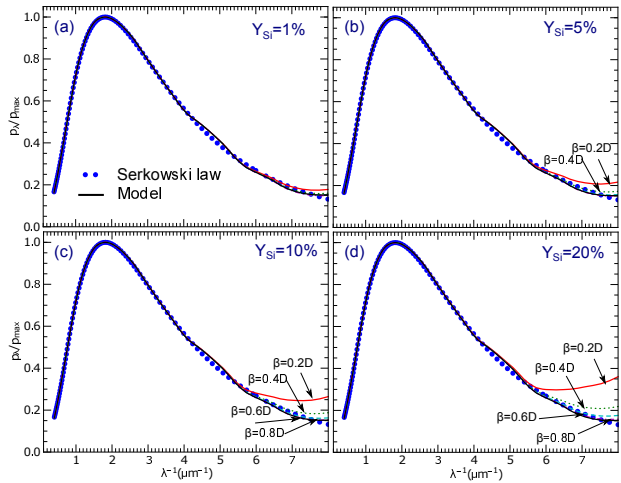}
\caption{Polarization of starlight computed for different values of $Y_{\rm Si}$ and $\beta$. Symbols show the observed polarization curve described by the Serkowski law with $\lambda_{\max}=0.55\mum$. The cases with $\beta\ge 0.4$D and $Y_{\rm Si}<10$ do not violet the UV polarization of $\lambda^{-1}\ge 6\mum^{-1}$.}
\label{fig:polUV}
\end{figure}

The obtained results are shown in Figure \ref{fig:polUV}, where the symbols show the Serkowski law. The UV polarization tends to increase with increasing $Y_{Si}$, as expected. The case with $\beta<0.2D$ easily exceeds the UV polarization for abundance $Y_{Si}\ge 5\%$, which can be ruled out. Moreover, the case of larger $\beta$ ( $\beta\ge 0.4$D) and $Y_{Si}\le 10\%$ have small impact on the increase in the UV polarization.

\begin{figure}
\includegraphics[width=0.4\textwidth]{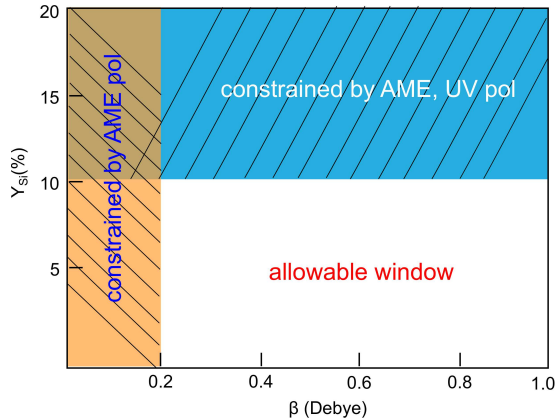}
\caption{Schematic of allowable parameters $\beta, Y_{\rm Si}$ for silicate nanoparticles for which its rotational emission and polarization do not violate observational constraints (white area). The orange shaded region represents the parameter space where the polarization violate the observed AME polarization, and the blue shaded area depicts the space where rotational emission violates the AME, and the polarization by nano silicate violates the UV polarization of starlight for the ISM.}
\label{fig:spinsil}
\end{figure}

Parameter space for spinning silicate that does not violate current observational constraints is sketched in Figure \ref{fig:spinsil}. It is worth mentioning that the constraint $Y_{\rm Si}< 10\%$ is consistent with with the estimate based on the UV extinction and mid-IR emission by \cite{Li:2001p4761}.

\subsection{Comparison with previous spinning dust studies}
Previous spinning dust models (DL98; \citealt{2009MNRAS.395.1055A}; \citealt{Silsbee:2011p5567}; HDL10; HLD11) dealt with spinning PAHs, because it is a well-established dust population in the ISM (see \citealt{2008ARA&A..46..289T}). Since physics of spinning silicate (iron) is essentially analogous to that of spinning PAHs, it is natural to take our advanced spinning dust model for modeling emission from spinning non-PAHs. 

It is noted that the last model (HDL10; HLD11) treats all known important physical effects, including grain wobbling due to thermal fluctuations within the grain, transient spin-up by single-ion collisions. In particular, the model can be applied to spinning dust grains of arbitrary shape and arbitrary temperature (HLD11). 

Finally, the spinning dust model (HDL10; HLD11) is the unique model that can predict self-consistently the polarization level of spinning dust emission (see also \citealt{{2013ApJ...779..152H},{2014ApJ...790....6H}}); the spinning model from other groups only provides rotational emissivity. It turns out that our model can be used to constrain the dipole moment and magnetic properties of nanoparticles using the AME polarization data. 

\section{Summary}\label{sec:sum}
In summary, in this paper, we have obtained the principal results as follows:

\begin{itemize}
\item[1] We have quantified rotational emission from spinning ultrasmall silicate grains for different grain size distribution parameters, various fraction of total Si abundance contained in ultrasmall grains $Y_{\rm Si}$, and a wide range of the dipole moment per atoms $\beta$.

\item[2] The polarization fraction of spinning silicate emission is found to increase with decreasing the value of $\beta$. The polarization fraction is below $5\%$ for the dipole moments $\beta>0.4$D, but it can reach $\sim 20\%$ for $\beta\sim 0.1$D.

\item[3] We identified that the presence of silicate nanoparticles with $Y_{Si}\le 10\%$ and the dipole moment $\beta\ge 0.4D$ does not violate the UV polarization of starlight and the observed polarization of AME. With this realization ($Y_{Si}\le 10\%, \beta\ge 0.4D$), rotational emission from spinning nano silicate can adequately reproduce the AME from the ISM.

\item[4] Our quantitative results suggest that rotational emission from spinning silicate nanoparticles may be an additional source of the AME. Future observations are needed to search for silicate emission features of nano silicate from the regions with prominent AME.
\end{itemize}

There are several remaining questions on the problem of spinning nano silicates, including the dipole moment and the abundance of nanoparticles $Y_{Si}$. In particular, unlike PAHs that have been identified through mid-IR emission features, the signature of silicate nanoparticles is not yet clear. Further studies looking for signatures of silicate nanoparticles are necessary for better understanding of the AME. Lastly, since this paper is intended to explore the importance of rotational emission from spinning silicate, we have presented calculations only for a standard model of the ISM with typical physical parameters characterized by the CNM. More calculations for the different phases of the ISM and fitting to observational data will be presented in our future paper.

\acknowledgments{TH is grateful to Francois Boulanger, Bruce Draine, Brandon Hensley, Alex Lazarian, Peter Martin, Nguyen-Luong Quang, and Svitlana Zhukovska for fruitful discussions. TH thanks Peter Martin for suggesting the constraint by UV polarization. TH thanks Alex Lazarian for valuable comments on the manuscript and warm hospitality during which the paper is complete. NQL and NAV are supported in part by the Ministry of Education (MOE) Grant No. B2014-17-45.}


\bibliography{ms.bbl}

\appendix
\section{A. Rotational dynamic timescales}\label{apdx:FGcoeff}

The thermal angular velocity of a grain around its symmetry axis (of rotational energy $k_{\B}T_{\gas}$) is
\bea
\omega_{T}=\left(\frac{2k_{B}T_{\gas}}{I_{\|}} \right)^{1/2}\simeq 3.3\times 10^{5}a_{-5}^{-5/2}s^{2/3}\hat{\rho}^{-1}\hat{T}_{\gas}^{1/2} \s^{-1},\label{eq:omegaT}
\ena
where $\hat{\rho}=\rho/3\g\cm^{-3}$, and $\hat{T}_{\gas}=T_{\gas}/100\K$.

The characteristic damping times of an oblate spheroidal grain for rotation along the directions parallel and perpendicular to the grain symmetry axis $\ahat_{1}$ are respectively given by (\citealt{1997MNRAS.288..609L})
\bea
\tau_{\H,\|}=\frac{3I_{\|}}{4\sqrt{\pi}n_{\H}m_{\H}v_{\th}a_{2}^{4}\Gamma_{\|}},~~\tau_{\H,\perp}=\frac{3I_{\perp}}{4\sqrt{\pi}n_{\H}m_{\H}v_{\th}a_{2}^{4}\Gamma_{\perp}},\label{eq:tauHxy}
\ena
where $\tau_{\H,\|}\equiv\tau_{\H,z}, \tau_{\H,\perp}\equiv \tau_{\H,y}=\tau_{\H,x}$ with $z$ the grain symmetry axis, and $x$ and $y$ being the axes perpendicular to the symmetry axis. In the above equation, $n_{\H}$ is the gas density, $m_{\H}$ is the hydrogen mass, $v_{\th}=(2k_{\B}T_{\gas}/m_{\H})^{1/2}$ is the thermal speed of hydrogen. Above, $\Gamma_{\|}$ and $\Gamma_{\perp}$ are geometrical factors, which are given by  (\citealt{1993ApJ...418..287R}):
\bea
\Gamma_{\|}=\frac{3}{16}\left[3+4(1-e^2)g(e)-e^{-2}(1-(1-e^2)^2)g(e)\right],\label{eq:Gam_par}\\
\Gamma_{\perp}=\frac{3}{32}\left[7-e^2+(1-e^2)^{2}g(e)+(1-2e^2)(1+e^{-2}
[1-(1-e^2)^2)g(e)])\right],\label{eq:Gam_per}
\ena
where $e=\sqrt{1-s^{2}}$ and 
\bea
g_{e}=\frac{1}{2e}\ln\left(\frac{1+e}{1-e}\right).\label{eq:ge}
\ena

{The limiting values are $\Gamma_{\|}=\Gamma_{\perp}=1$ for $e=0$ (i.e., spherical grains), and $\Gamma_{\|}=\Gamma_{\perp}=3/8$ for $e=1$.}

For the typical parameters of the ISM, Equations (\ref{eq:tauHxy}) become
\bea
\tau_{\H,\|}\simeq 6.58\times 10^{4} \hat{\rho}\hat{s}^{2/3} a_{-5}
\hat{n}_{\H}^{-1}\hat{T}_{\gas}^{-1/2} \Gamma_{\|}^{-1}\yr,~~~
\tau_{\H,\perp}\simeq 4.11\times 10^{4}\hat{\rho}\hat{s}^{2/3}\left(\frac{1+s^{2}}{1.25}\right) a_{-5}\hat{n}_{\H}^{-1}\hat{T}_{\gas}^{-1/2} \Gamma_{\perp}^{-1}\yr,
\ena
where $\hat{s}=s/0.5$.

Likewise, the characteristic damping times due to the rotational dipole emission can be rewritten as (see HDL10):
\bea
\tau_{\ed,\|}=\frac{3I_{\|}c^{3}}{6k_{\B}T_{\gas}\mu_{\perp}^{2}},~~
\tau_{\ed,\perp}=\frac{3I_{\perp}c^{3}}{6k_{\B}T_{\gas}\left(\mu_{\perp}^{2}/2+\mu_{\|}^{2}\right)},\label{eq:tauedxy}
\ena
where $\mu_{\|}$ and $\mu_{\perp}$ are the components of the dipole moment $\bmu$ parallel and perpendicular to the grain symmetry axis.

\begin{figure*}
\centering
\includegraphics[width=0.3\textwidth]{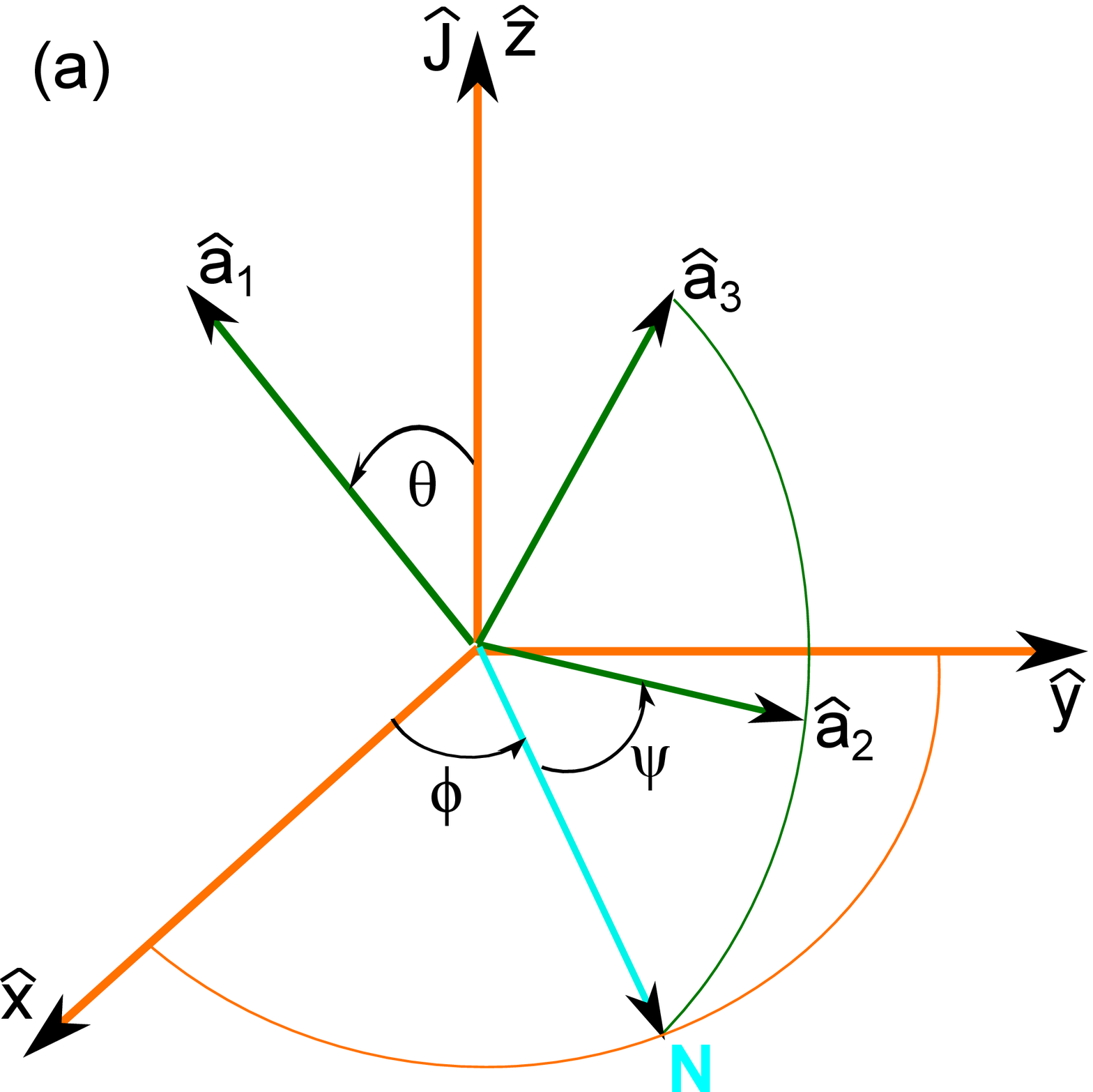}
\includegraphics[width=0.35\textwidth]{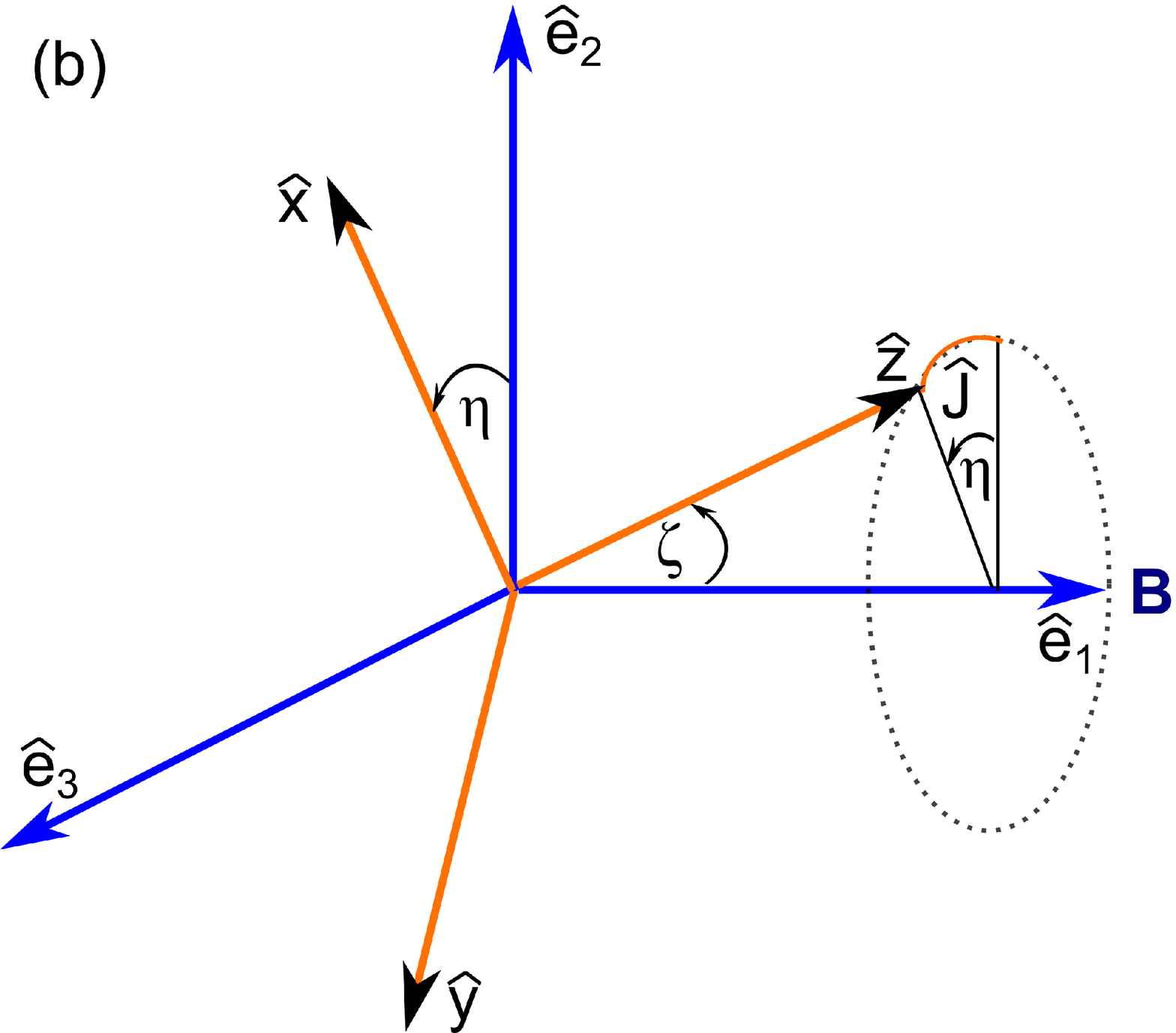}
\caption{Coordinate systems used for calculations. (a) Orientation of grain principal axes in the coordinate system $\xhat\yhat\zhat$ with $\zhat$ parallel to the grain angular momentum $\bJ$. (b) Orientation of $\bJ$ in the inertial coordinate system $\ehat_{1}\ehat_{2}\ehat_{3}$ with $\ehat_{1}$ parallel to the magnetic field $\Bv$.}
\label{fig:RFs}
\end{figure*}
\section{B. Magnetic Alignment of thermally rotating grains}\label{sec:DG}
Since we are interested also in polarization of spinning dust emission, it is worth to briefly summarize the magnetic property of silicate material and magnetic alignment.

Silicate material is an ordinary paramagnetic material. The zero-frequency susceptibility is given by
\bea
\chi(0)=0.042f_{p}\left(\frac{15K}{T_{d}}\right),
\ena
where $T_{d}$ is the grain temperature and $f_{p}$ is the fraction of paramagnetic atoms (i.e., atoms with partially filled shells) in the grain (see Draine 1996 and references therein). For presented calculations, a typical value $f_{p}=0.1$ is assumed for the ordinary paramagnetic material of silicate.

The frequency-dependence imaginary part of susceptibility $\chi_{2}(\omega)$ is given by the critically-damped solution:
\bea
\chi_{2}(\omega)&=&\frac{\chi(0) \omega \tau}{[1+(\omega \tau/2)^{2}]^{2}}\label{eq:chi_cd},
\ena
where $\tau=\tau_{2}\sim 2.9\times 10^{-12}/f_{p}\s$ is the spin-spin relaxation time (see \citealt{2014ApJ...790....6H}).

\cite{1951ApJ...114..206D} (henceforth DG51) suggested that a paramagnetic grain rotating with angular velocity $\bomega$ in an external magnetic field experiences paramagnetic relaxation, which dissipates the grain rotational energy into heat. This results in the gradual alignment of $\bomega$ and $\bJ$ with the magnetic field at which the rotational energy is minimum.

The characteristic timescale for the magnetic alignment of $\bJ$ with $\bB$ is given by
\bea
\tau_{\rm m} =\frac{I_{\|}}{K(\omega)VB^{2}}=\frac{2\rho a^{2}s^{-2/3}}{5K(\omega)B^{2}},\label{eq:tau_DG}
\ena
where $K(\omega)=\chi_{2}(\omega)/\omega$, and $I_{\|}$ from Equation (\ref{eq:Iparperp}) has been used. 

To describe the effect of grain alignment by magnetic relaxation against the randomization by gas atoms, a dimensionless parameter $\delta_{m}$ is usually used:
\bea
\delta_{m}=\frac{\tau_{\H,\|}}{\tau_{\rm m}}\simeq 0.28\frac{ a_{-5}^{-1}s^{4/3}\hat{B}^{2}(K(\omega)/10^{-13}\s^{-1})}{\hat{n}_{\gas}\hat{T}_{\gas}\Gamma_{\|}},\label{dis3}
\ena
where $\tau_{\H,\|}$ is the gaseous damping for the rotation around the grain symmetry axis, $\hat{n}_{\gas}=n_{\H}/30\cm^{-3}$, $\hat{T}_{\gas}=T_{\gas}/100\K$, $\hat{B}=B/10\mu{\rm G}$, and $\Gamma_{\|}<1$ is a geometrical factor (see Appendix \ref{apdx:FGcoeff}). 

The fluctuation terms associated with the magnetic dissipation are  
\bea
B_{m,xx}=B_{m,yy}=\frac{T_{d}}{T_{gas}}\delta_{m},\label{eq:Bmxy}
\ena
and $B_{m,zz}=0$.

\end{document}